# Towards Architectural Programming of Embedded Systems

Arne Haber, Jan O. Ringert, Bernhard Rumpe
Software Engineering,
RWTH Aachen University, Germany
http://www.se-rwth.de/

**Abstract:** Integrating architectural elements with a modern programming language is essential to ensure a smooth combination of architectural design and programming. In this position statement, we motivate a combination of architectural description for distributed, asynchronously communicating systems and Java as an example for such an integration. The result is an ordinary programming language, that exhibits architecture, data structure and behavior within one view. Mappings or tracing between different views is unnecessary. A prototypical implementation of a compiler demonstrates the possibilities and challenges of architectural programming.

## 1 Java with Architectural Elements

As stated in [MT00] there are a number of languages that support design, analysis, and further development of software-system-architectures. These languages are commonly known as Architecture Description Languages (*ADL*) and allow a high level description of software systems of a specific domain. Using an ADL enables reasoning about specific system properties in an early development stage [GMW97]. Furthermore, there are quite often mappings from architecture to a General Purpose Language (*GPL*), producing code frames for the GPL. This helps ensuring the architectural consistency initially, but when the code evolves the architecture becomes implicitly polluted or when the architecture shall be evolved this needs to be done on the code level. Tracing is therefore important to keep architecture and code aligned. However, it would be much better to integrate both, architecture and code into one single artifact such that tracing is not necessary anymore. [MT00] defines a component as a unit of computation or storage that may represent the whole software system or just a single small procedure. Components in distributed systems partially run in a distributed manner on different hardware. As a consequence they do not share memory and the communication through shared variables or ordinary method calls is not feasible. So they communicate with each other through channels called connectors by asynchronous message passing. Here a component always has an explicitly defined interface of needed and offered connectors. In contrast to this, object oriented classes respectively their instances in a GPL are mostly accessed through their implemented interfaces synchronized by blocking method calls. But a class does not explicitly describe the interfaces it uses. In addition, the hierarchical containment that is intensely used in ADLs to structure systems, is almost completely missing in the object oriented paradigm. A common way to structure object oriented systems is the usage of packages. However

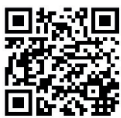

one is not able to express hierarchical containment with this technique.

Several approaches like JavaBeans [JB09] enrich an existing GPL with a component concept. Nevertheless they do not exceed or extend the borders drawn by the target object oriented GPL. These approaches mainly introduce new libraries written in plain GPL code or map a language extension into the original GPL. Doing so, Java has been enriched by JavaBeans with a concept to use components in an object oriented environment, but the traceability from architecture to code has not been increased very much. And this traceability is necessary, because the developer is exposed with both views, the architecture and the code.

We believe that one way to raise this traceability respectively to make it unnecessary is to combine an existing ADL with a common GPL in such a way that architectural definitions are explicit and essential part of the language. We decided to use the ADL MontiArc that resembles a common understanding of how to model distributed communicating systems, similar to automotive function nets [GHK$^+$07], or UML's composite structure diagrams [OMG07], and with a precise underlying calculus like FOCUS [BS01] as described in Sect. 4. As our target GPL we decided to use Java, because it is a widely accepted modern language. We integrate classes, methods, and attributes into components. This gives us a new programming language with the working title "AJava" and enables us to combine concrete behavior descriptions with an abstract high-level architectural description directly in one artifact. Enhanced with a syntax highlighting Eclipse-editor that supports functions like auto-completion, folding, error messages, and hyperlinking to declarations, one is able to program architectures in a familiar and comfortable development environment. Further tool support is given by a prototypical compiler for a subset of AJava based on the DSL framework MontiCore [GKR$^+$08].

The concrete syntax of AJava will be shown in the next section with an introducing example. In Sect. 3 we discuss aspects of the design and variations of AJava. Our approaches to building a compiler and defining semantics are presented in Sect. 4. This paper ends with related approaches and a conclusion in sections 5 and 6.

## 2 Integrated Component Programming

As an example of how to model and implement an embedded system in AJava we present a coffee machine which takes the selection of a type of coffee as input. Connected to the machine is a milk dispenser which is managed by the coffee machine specifying the needed amount of milk and receiving an error signal if the milk tank is empty. The coffee machine itself is composed of a display, the coffee processing unit and bean sensors to monitor the available amount of coffee and espresso beans.

A graphical representation of the main component is given in Fig. 1. Its corresponding implementation in AJava is given in listing 1. Please note that the textual representation is covering all features in Fig. 1, although some connectors are not explicitly given but derived from the component's context. We assume, regarding the development process, that appropriate editors show the textual representation as main development artifact and the diagram as outline for a better navigation. The component `CoffeeMachine` in listing 1 defines a public Java enumeration `CoffeeType` (ll. 8–9) that can be used by all other components. The interface of a component (its ports) is declared after the keyword `port` (ll. 3–6) where each port needs a type `T`. Admissible types are all Java types. The port then accepts incoming data of type `T` or its subtypes. If port names are omitted they will be set to their type's name (if it is unique).

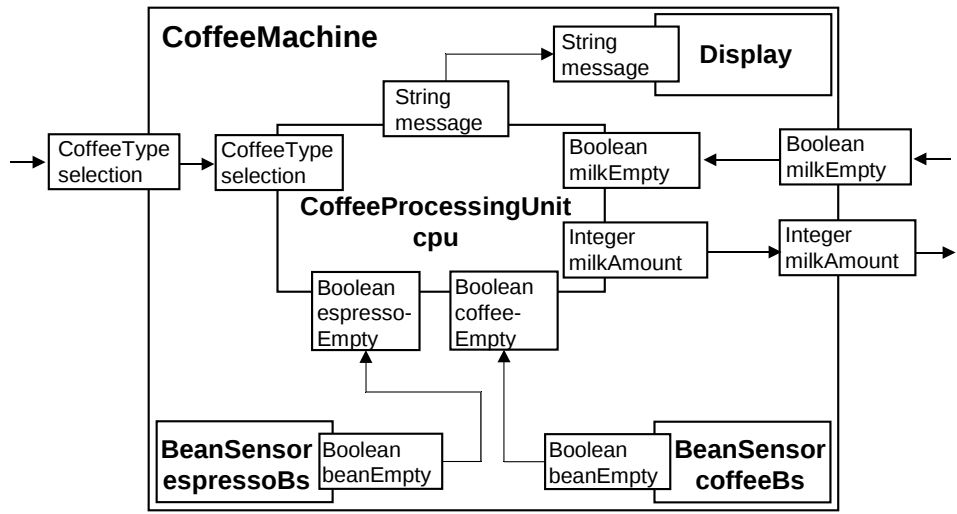

Figure 1: Architecture of the `CoffeeMachine` component

Inner components are instantiated using the keyword `component`. E.g. in l. 15 a component of type `Display` is instantiated and in l. 14 a component of type `CoffeeProcessingUnit` is created with name `cpu`. Naming components is optional as long as types are unique. Further elements of the architectural description are connectors that connect one outgoing port with one to arbitrarily many incoming ports. Explicit connections are made via the `connect` statement (l. 17) or at instantiation of an inner component. This short form for connectors can be seen in l. 12 connecting port `beanEmpty` of inner component `espressoBS` to a corresponding port of the coffee processing unit. In many cases explicit definition of connectors is not necessary if connections can be derived by port names. The automatic derivation is activated by `autoconnect port` (l. 2) but can be overridden and extended by explicit connections. Messages can only be transmitted (asynchronously) between ports via these communication channels, there is no other form of inter-component communication (cf. Sect. 3).

The behavioral component `CoffeeProcessingUnit` (*CPU*) is displayed in listing 2. In contrast to architectural components like `CoffeeMachine` behavioral components contain no further components but implement a concrete behavior (cf. Sect. 3.3). The `CoffeeProcessingUnit` contains an interface declaration (ll. 2–8) like all components in AJava do. The CPU declares a private state variable `milkAvailable` (l. 10) and amongst others the dispatcher method `onMilkEmptyReceived` (l. 12). This

```
component CoffeeMachine {
  autoconnect port;
  port
    in CoffeeType selection,
    in Boolean milkEmpty,
    out Integer milkAmount;

  public enum CoffeeType
    { LatteMacchiato, Espresso, Cappucino, Coffee }

  component BeanSensor
    espressoBS [beanEmpty->cpu.espressoEmpty],
    coffeeBS;
  component CoffeeProcessingUnit cpu;
  component Display;

  connect coffeeBS.beanEmpty->cpu.coffeeEmpty;
}
```

Listing 1: Structural component `CoffeeMachine` in AJava syntax

method by convention dispatches incoming messages arriving on port `milkEmpty` of type `Boolean`. Thus the communication is event triggered, but other implementations will be possible, where data is buffered by the environment or the component's ports, allowing a possibly explicitly defined scheduling strategy to manage the input buffer. The example method (ll. 12–19) reacts on input from a sensor and sends an according text message via its outgoing port `message` (ll. 14, 16). This port is connected to the display of the coffee machine (cf. Fig. 1) which is not known inside the CPU component. Please note that outgoing ports have similarities to private variables and in their implementation they offer a sending procedure to transmit data from this port.

## 3 Discussion of the Designed Language

The proposed language AJava is pointing out one way towards a new paradigm or at least a paradigm integration between object-orientation and architectural design. The resulting language will not be a silver bullet, but should enable programmers to write evolvable and well-structured code more easily. However, many language design decisions are still to be considered based on empirical evidence that is collected using this prototypical language.

In the following, some issues on the design of AJava and its semantics are introduced and discussed. They mostly tackle trade-offs between Java's features for interaction between objects, and harnessing the complexity of the architecture described in AJava.

```
 1  component CoffeeProcessingUnit {
 2    port
 3      in CoffeeType selection,
 4      in Boolean espressoEmpty,
 5      in Boolean coffeeEmpty,
 6      in Boolean milkEmpty,
 7      out Integer milkAmount,
 8      out String message;
 9
10    private boolean milkAvailable;
11    //...
12    public void onMilkEmptyReceived(Boolean milkEmpty) {
13      if (milkEmpty) {
14        this.message.send("Sorry, no milk today.");
15      } else {
16        this.message.send("Got milk!");
17      }
18      this.milkAvailable = !milkEmpty;
19    }
20  }
```

Listing 2: The coffee processing unit implemented in AJava

### 3.1 Communication forms between components

Components in AJava can contain attributes, variables, and even encapsulated Java classes. Intuitively components are similar to objects and could use their regular ways of communication. This would allow method calls, event passing, shared variables and more mechanisms used in object oriented systems to also be used among components. The design of ADLs and MontiArc in general favors however a more limited form of communication: message passing via channels connecting ports of components. In the context of AJava this restriction would prohibit that a component calls a method or reads attributes of another component. The second communication form especially ensures the classical communication integrity (cf. [LV95]) that claims that components only communicate through explicitly defined interfaces and such the effect and resulting behavior of a reused component is much easier to predict.

### 3.2 Communication via Channels

Channels are typed connections from one source port to one or more target ports. Channels can also be fed back, where source and target port belong to the same component. A programmer might want to pass references to objects from one component to another to share object state. While this might be convenient and feasible if two components run in the same VM, it pollutes clean design, because it forces components to run on the same physical engine. Furthermore, it couples components in an implicit way, making reuse much more difficult. Language design variants are to only communicate via (a) messages of simple types or (b) encode the full object structure and send it over the channel. The latter however, can lead to much unnecessary traffic and might lead to semantic changes due to different object identities of otherwise identical objects. This could be improved through optimizing strategies, e.g. a transparent implementation of lazy sending of data parts, and explicit encoding of object identities.

### 3.3 Structural vs. Behavioral Components

Several ADLs like ROOM [SGW94] force the system designer to compose system-behavior as the accumulated behavior of all leaf-components in the lowest hierarchy layer. Other approaches like [OL06], in that case UML Composite Structure Diagrams are used to model architectures, allow behavior on all hierarchy-layers of system-architectures. On the one hand, both variants can be translated into each other. On the other hand, reuse and convenient modeling of parts of the software seem to be pretty much affected by these two different approaches.

For example a structural refinement does not necessary yield to a behavioral refinement in the latter case. By now AJava follows the first strategy and separates between structural and behavioral components. We believe that the effort needed to break down functionality to leaf-nodes pays off with better traceability of functionality and the ability to replace whole branches with dummies for better testability.

However, experiments of a controlled mixture of behavioral elements and a structural decomposition within a component could show, how e.g. to integrate scheduling or message dispatching strategies within components that allow the programmer a fine grained control over the messages processed.

### 3.4 Embedded components

In contrast to general purpose components running on regular VMs an embedded AJava component should be able to run on an embedded systems with few resources. Compiling against small VMs like kaffe [Kaf09] or JamVM [Jam09] restricts the used Java version and compiler. JamVM is extremely small but only supports the Java Language Specification Version 2, so some new concepts like e.g. generics or annotations are not available. To avoid this drawback we might use Java SE For Embedded Use [EJ09] that is currently compatible with Java 6. However this VM requires much more resources and reduces the application range of AJava components to devices with more powerful cpus like e.g. mobile phones. Please note that AJava can, besides its application for embedded systems, also be used for general purpose software development tasks.

# 4 Language Realization and Semantics

For a precise understanding of the AJava language, a formal specification of the key concepts is most helpful. For a definition of its features and semantics we follow the methodology proposed in [HR04]. In a first step we defined its syntax as a context free grammar using the DSL framework MontiCore [GKR+08], a framework for quick development of modeling and also programming languages. From there we explored and still explore the language through application of a prototypical AJava compiler. The objective of this compiler is to translate AJava sources to complete and self-contained Java classes. The system engineer only works on AJava artifacts and not the generated Java code. Although as sketch of the formal denotational semantics is understood, a precise definition will later be defined based on [GRR09] to define system model semantics for AJava.

The current implementation of the MontiCore compiler generator derives the abstract syntax tree as well as an instance of the ANTLR parser [Ant08] that can process AJava programs. In previous works MontiCore grammars for Java as well as an architectural description language MontiArc have been developed independently. As MontiCore supports the reuse and combination of independently defined languages e.g. through embedding of languages [KRV08b] the development of a composed compiler was relatively quick and straightforward.

## 4.1 Communication realization

As discussed in Sect. 3 several variants of communication mechanisms are possible, scaling from strict port communication to liberal method calls between any kinds of objects. While method calls can be realized directly in Java, port to port communication in a distributed system has to be implemented in a different way. Ideally components need not care about the physical deployment of their communication partners. This also means that all components, either on one machine or distributed, use the same communication interface. Generally components run in their own thread and asynchronous communication is realized through buffering to decouple components.

This creates the need for a smart communication layer encapsulating the inter-component communication. As AJava components are realized in Java, existing communication methods like RMI [RMI09] or CORBA [OMG08] are of particular interest. As both protocols use TCP for inter-component communication they are not feasible for embedded bus communication. Instead a hand written communication layer that maps component port communication to e.g. the Java CAN API [BN00] is more suitable. Additional capabilities, like buffering or an explicit possibility of a component to manipulate its input buffer, will need extra infrastructure to be implemented. The suitability of different communication layers for AJava components in an embedded environment is to be investigated.

### 4.2 Formal Semantics

Operational semantics of AJava is defined by supplying a translation to Java code. This definition of semantics makes it hard to apply automated reasoning about features of the language and properties of programs since the translation rules are not explicitly given in a way accessible to theorem provers. So far there is no complete reasoning framework for the whole Java language. Advances however have been made in formalizing the language, and proving the type system correct [Ohe01].

The approach favored here is the semantic mapping of AJava to the system model described in [BCR06, BCR07a, BCR07b]. This approach has been introduced in [CGR08a, CGR08b] for class diagrams and state charts. Inter-object communication can be specified in two ways in the system model: composable state machines or through communication of components in FOCUS style [BCR07b].

A logic for realizable component networks based on [GR07] is currently under development and will be suited for the definition of AJava's semantics with respect to certain of the above discussed design decisions. For the most natural semantics component communication directly maps to asynchronous communication over typed Focus channels. Both approaches mentioned so far only support the abstract definitions of timing. Reasoning about timing in real-time systems thus has to be further investigated.

## 5 Related Work

Another approach to combine the GPL Java with an ADL is ArchJava [ACN02] that introduces components, ports, and connectors to an object oriented environment. In contrast to AJava ArchJava facilitates dynamic architectures as described in [MT00]. This way components can be added and removed to systems during runtime. The major disadvantage of ArchJava compared to AJava is that ArchJava's components have no support of distribution across the borders of a single VM.

A common approach to implement components in Java is the usage of JavaBeans [JB09]. Again, the main disadvantage compared to an integrated ADL based programming language is either a missing mapping from architecture to code or the need for synchronization of many different heterogeneous documents describing the system. In addition JavaBeans communicate via method calls hence they are bound to run on a single VM. To avoid this drawback they can be combined with RMI, but as discussed in Sect. 4.1, this is not feasible for embedded systems.

Another approach to ensure architectural consistency is presented in [DH09]. The described prototype ArchCheck automatically derives a base of logical facts from java source code describing existing dependencies. Architectural consistency rules are manually derived from architectural descriptions and checked by an interpreter according to the derived base of rules and facts. Compared to AJava this approach is not bounded to one system domain as AJava only copes information flow architectures. But again, the involved artifacts have to be synchronized manually in contrast to AJava that automatically enforces the architectural consistency by design of the language.

## 6 Conclusion

We propose a possible way to combine architectural and concrete behavioral descriptions into the programming language AJava. This language integrates Java and the ADL MontiArc rendering artificial mappings between architecture and implementation superfluous.

This work is still in a preliminary stage. Based on our tooling framework [GKR+08, KRV08b, KRV08a], we are currently developing an enhanced version of the compiler for our language.

This prototype together with other experience on definition of programming languages [Rum95] will help us to contribute to a possible smooth extension of a GPL with appropriate architectural elements, such that the level of programming is raised towards architecture and may be in the future also towards requirements.